\documentclass[twocolumn]{aastex62}

\graphicspath{{./}{}}

\received{June 29, 2019}
\revised{September 28, 2019}
\accepted{October 4, 2019}
\submitjournal{ApJL}

\shorttitle{HSC-XD 52}
\shortauthors{Halevi et al.}

\begin{document}

\title{HSC-XD 52: An X-ray detected AGN in a low-mass galaxy at $z\sim0.56$}

\correspondingauthor{Goni Halevi}
\email{ghalevi@princeton.edu}

\author[0000-0002-7232-101X]{Goni Halevi}
\affil{Department of Astrophysical Sciences, Princeton University, 
4 Ivy Lane, Princeton, NJ 08544, USA}

\author{Andy Goulding}
\affil{Department of Astrophysical Sciences, Princeton University,
4 Ivy Lane, Princeton, NJ 08544, USA}

\author{Jenny Greene}
\affil{Department of Astrophysical Sciences, Princeton University,
4 Ivy Lane, Princeton, NJ 08544, USA}

\author{Jean Coupon}
\affil{Department of Astronomy, University of Geneva,
ch.\ d'Ecogia 16, 1290, Versoix, Switzerland}

\author{Anneya Golob}
\affil{Department of Astronomy \& Physics, Institute for Computational Astrophysics, Saint Mary's University, Halifax, Canada}

\author{Stephen Gwyn}
\affil{NRC Herzberg Astronomy and Astrophysics, 5071 West Saanich Road, Victoria, BC V9E 2E7, Canada}

\author[0000-0001-9487-8583]{Sean D. Johnson}
\altaffiliation{Hubble \& Carnegie-Princeton fellow}
\affil{Department of Astrophysical Sciences, Princeton University, 
4 Ivy Lane, Princeton, NJ 08544, USA}
\affil{The Observatories of the Carnegie Institution for Science, 813 Santa Barbara Street, Pasadena, CA 91101, USA}

\author[0000-0002-3305-9901]{Thibaud Moutard}
\affil{Department of Astronomy \& Physics, Institute for Computational Astrophysics, Saint Mary's University, Halifax, Canada}

\author[0000-0002-7712-7857]{Marcin Sawicki}
\altaffiliation{Canada Research Chair}
\affil{Department of Astronomy \& Physics, Institute for Computational Astrophysics, Saint Mary's University, Halifax, Canada}
\affil{NRC Herzberg Astronomy and Astrophysics, 5071 West Saanich Road, Victoria, BC V9E 2E7, Canada}

\author{Hyewon Suh}
\altaffiliation{Subaru Fellow}
\affil{Subaru Telescope, National Astronomical Observatory of Japan,
650 North A'ohoku Place, Hilo, HI 96720, USA}

\author{Yoshiki Toba}
\affil{Department of Astronomy, Kyoto University,
Kitashirakawa-Oiwake-cho, Sakyo-ku, Kyoto 606-8502, Japan}
\affil{Academia Sinica Institute of Astronomy and Astrophysics, 11F of Astronomy-Mathematics Building, AS/NTU, No.1, Section 4, Roosevelt Road, Taipei 10617, Taiwan}
\affil{Research Center for Space and Cosmic Evolution, Ehime University, 2-5 Bunkyo-cho, Matsuyama, Ehime 790-8577, Japan}

\begin{abstract}
The properties of low-mass galaxies hosting central black holes provide clues about the formation and evolution of the progenitors of supermassive black holes. In this letter, we present HSC-XD 52, a spectroscopically confirmed low-mass active galactic nucleus (AGN) at an intermediate redshift of $z\sim0.56$. We detect this object as a very luminous X-ray source coincident with a galaxy observed by the Hyper Suprime-Cam (HSC) as part of a broader search for low-mass AGN. We constrain its stellar mass through spectral energy distribution modeling to be LMC-like at $M_\star \approx 3 \times 10^9 M_\odot$, placing it in the dwarf regime. We estimate a central black hole mass of $M_\mathrm{BH} \sim 10^{6} M_\odot$. With an average X-ray luminosity of $L_X \approx 3.5 \times 10^{43}~\mathrm{erg}~\mathrm{s}^{-1}$, HSC-XD 52 is among the most luminous X-ray selected AGN in dwarf galaxies. The spectroscopic and photometric properties of HSC-XD 52 indicate that it is an intermediate redshift counterpart to local low-mass AGN.

\end{abstract}



\section{Introduction} \label{sec:intro}

Supermassive black holes (SMBHs) are thought to play a crucial role in galaxy evolution. SMBHs are ubiquitous in present-day massive galaxies and their properties correlate with those of their hosts \citep[for a review, see][]{2013ARA&A..51..511K}. Observations of central black holes in dwarf galaxies over cosmic time may provide insight into the birth and growth of SMBHs. Simulations predict that these low-mass black holes experience relatively little growth via mergers or accretion \citep{2011ApJ...742...13B}, allowing them to serve as indirect probes of the black holes that seed the SMBHs observed in massive galaxies today.

At present, we know little about the demographics of central black holes in low-mass galaxies, even in the local universe. The dynamical signatures of $\sim 10^5~M_\odot$ massive black holes (MBH) are found in some, but not all, galaxies within $3.5$~Mpc with  $M_\star \approx 10^9-10^{10} M_\odot$ \citep{2019ApJ...872..104N}.
The dearth of information about black holes in this regime not only limits our understanding of SMBH seeds but also impedes predictions of gravitational wave events detectable by LISA and their rates \citep[e.g.,][]{2019MNRAS.482.2913B}.

Statistical constraints on the black hole occupation fraction require a large survey of black holes in low-mass ($M_\star \lesssim 10^{10} M_\odot$) hosts. Searches with optical emission lines \citep[e.g.,][]{1997ApJS..112..315H,2007ApJ...670...92G,2012ApJ...755..167D,2013ApJ...775..116R,2014AJ....148..136M}, mid-infrared (IR) spectroscopy \citep[e.g.,][]{2009ApJ...704..439S}, and radio continuum \citep[e.g.,][]{2014ApJ...787L..30R} have all yielded interesting samples. We focus on X-ray searches, which provide an unbiased measure of the accretion luminosity, and are relatively insensitive to obscuration \citep[e.g.,][]{2015A&ARv..23....1B}.

X-ray observations are a powerful tool to study MBH demographics locally \citep[e.g.,][]{2009ApJ...690..267D,2009ApJ...700.1759G,2011ApJ...728...25M,2012ApJ...753...38S,2012ApJ...757..179A,2015ApJ...805...12L,2017ApJ...842..131S}; they place the only accretion-based constraint on the occupation 
fraction \citep[$>20\%$;][]{2015ApJ...799...98M}. Pushing to higher redshift is enabled by ongoing deep X-ray observations \citep[e.g.,][]{2013ApJ...773..150S,2014ApJ...782...22B}, but it is very challenging to understand the completeness of spectroscopic samples \citep{2016ApJ...831..203P} and the purity of photometric samples \citep{2018MNRAS.478.2576M}.

In this letter, we present one tantalizing object from our new search for such sources, which uses the relatively wide survey area and sensitivity of the Deep Layer of the Hyper Suprime-Cam Subaru Strategic Program \citep[HSC-D;][]{2018PASJ...70S...8A} and the complementary CFHT Large Area U-band Deep Survey \citep[CLAUDS;][]{Sawicki2019} to find faint low-mass galaxies (G.~Halevi et al. 2019, in preparation). We conducted our search in the XMM-Newton Large-Scale Structure (XMM-LSS) field, where we used the {\sc SExtractor}-based $u^*grizy$ catalog produced by the CLAUDS team \citep[A.~Golob et al., in preparation; see also \S~3.1.2 in ][]{Sawicki2019}. This catalog extends the HSC photometric sampling, enabling improved estimates of the redshift, stellar mass ($M_\star$), and star formation rate (SFR) for each source (T.~Moutard et al., in preparation). To identify candidates for our sample of HSC X-ray Dwarfs (HSC-XDs), we cross-matched candidate dwarf galaxies in this catalog with the XMM-SERVS source catalog \citep{2018MNRAS.478.2132C}.

The source presented in this letter, hereafter referred to as HSC-XD 52, is an intermediate redshift low-mass AGN observed at multiple epochs in both X-ray imaging and optical spectroscopy. The galaxy has a dwarf-like stellar mass and hosts a luminous X-ray detected AGN which is seen to be declining in activity with time. Its position, basic properties, and derived best-fit parameters are provided in Table \ref{tab:props}.

\section{A luminous X-ray source in a dwarf galaxy} \label{sec:props}

\begin{figure}[htp]
    \centering
    \includegraphics[width=0.48\linewidth]{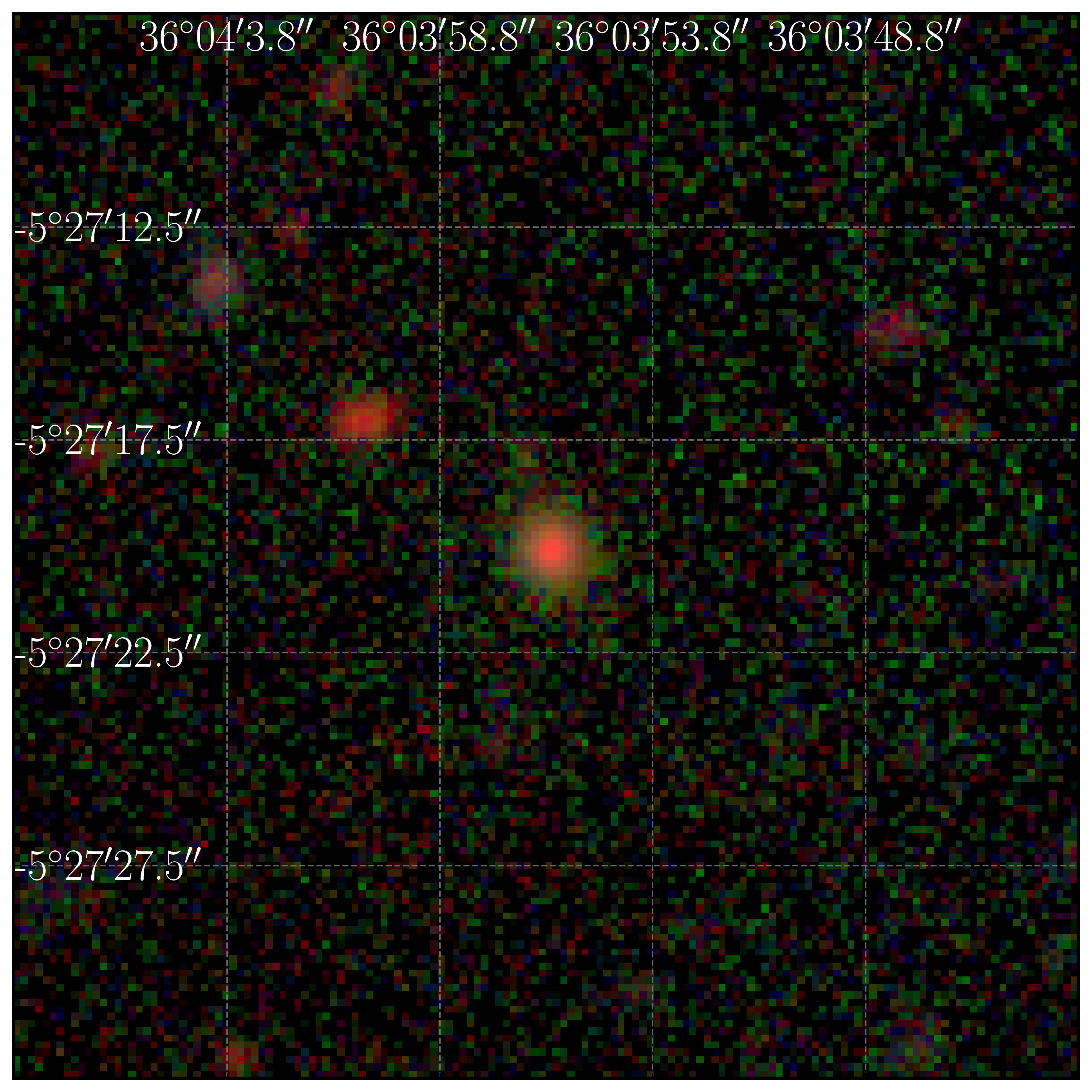} \includegraphics[width=0.48\linewidth]{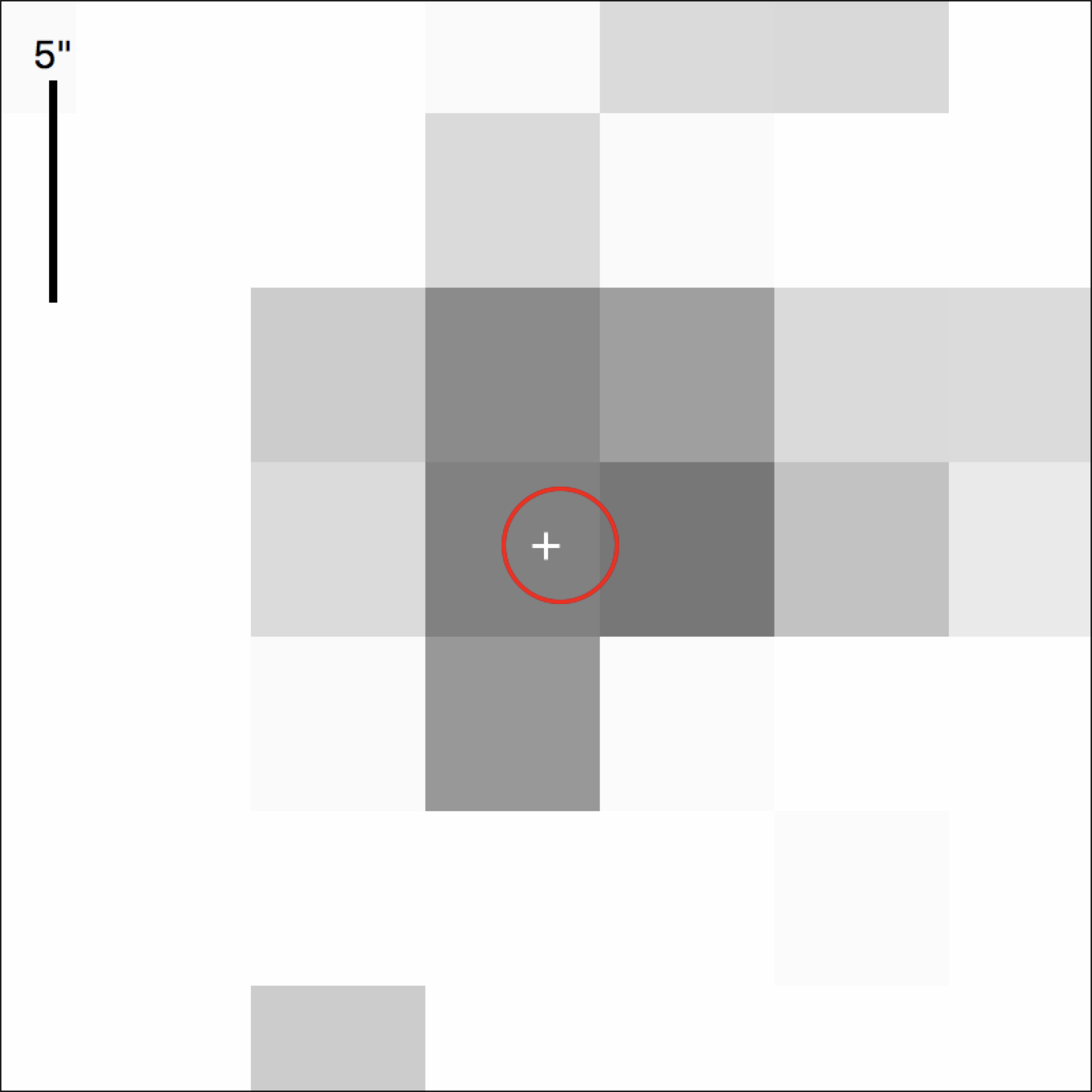}
    \includegraphics[width=\linewidth]{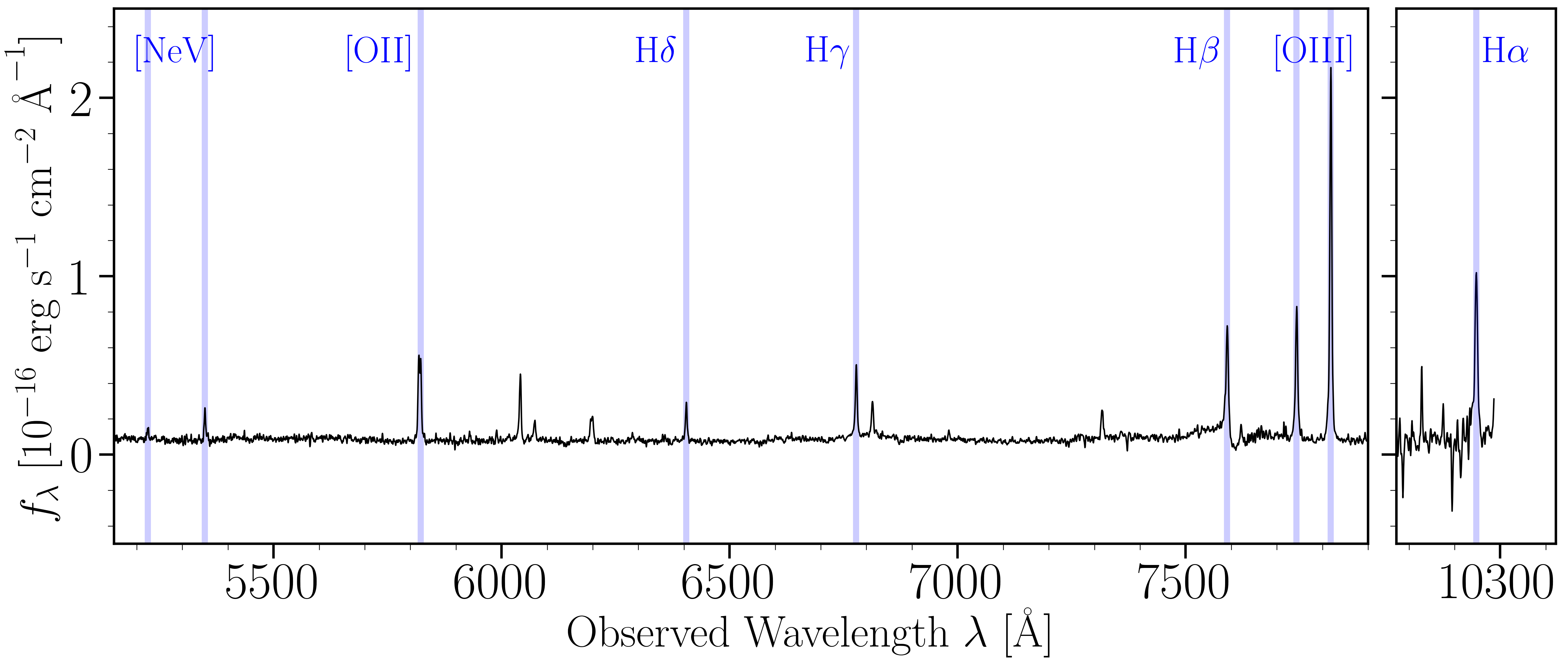}
    \caption{HSC-D $gri$ composite image (top left), XMM full-band image (top right), and smoothed MagE spectrum (bottom) of HSC-XD 52. The images span 25$''$ on each side. The overlaid grid has spacings of 5$''$. The XMM image uses an inverted logarithmic color scale. The white cross marks the XMM centroid, whereas the red circle with a radius of 1.3$''$ is centered on the HSC position. We note that the H$\beta$ emission line is blended with a strong sky line; its apparent broadness and asymmetry are not intrinsic.}
    \label{fig:ims+spec}
\end{figure}

HSC-XD 52 is of particular interest because the available data not only confirms its (time-evolving) AGN nature but also strongly implies a low stellar mass ($<10^{10} M_\odot$; assuming a Chabrier initial mass function). The HSC imaging shows a marginally extended red galaxy (top left panel of Fig. \ref{fig:ims+spec}; Table \ref{tab:props}). The XMM-SERVS data reveals a luminous X-ray source (top right panel of Fig. \ref{fig:ims+spec}) spatially coincident with the optically detected galaxy. Its X-ray properties are described in \S \ref{sec:BH}. The three epochs of XMM observations spanning 2006 to 2017 show temporal variability indicative of a fall in X-ray luminosity.

The Sloan Digital Sky Survey \citep[SDSS;][]{2017AJ....154...28B} obtained a spectrum of HSC-XD 52 despite its relative optical faintness ($i= 21.5$ mag) because it was targeted as part of the XMM-XXL follow-up program. We additionally acquired a spectrum on July 27 2019 with the Magellan Echellette (MagE) Spectrograph, a moderate-resolution ($R\sim4100$ for a $1''$ slit) optical echellette mounted on the Clay Magellan II telescope. While the SDSS spectrum from 2013 suggests the presence of broad lines indicative of accretion onto an MBH, this evidence is not present in the 2019 spectrum despite its higher resolution and signal-to-noise ratio (SNR). The broadband spectral energy distribution (SED) presented in Fig. \ref{fig:sed} and discussed in \S \ref{sec:Mstar} supports the classification of this source as an AGN in a dwarf galaxy, as do its early epoch X-ray properties and the ratios of its narrow emission lines. The combination of these properties points toward the characterization of HSC-XD 52 as a higher redshift analog of the $z\sim0$ low-mass AGN POX 52 \citep{2004ApJ...607...90B} and NGC 4395 \citep{2003ApJ...588L..13F}, though one with a accretion rate that is decreasing with time.

\begin{deluxetable}{cccc}
\tablecaption{Properties of HSC-XD 52 \label{tab:props}}
\tablehead{\colhead{$\alpha$} & \colhead{$\delta$} & \colhead{XMM ID} & \colhead{$z$}  
}
\startdata
02:24:15.76 &  -05:27:20.02 & XMM03471 & 0.561 \\ \hline \hline
\colhead{$i$ } & \colhead{$g-r$} & \colhead{$f_\mathrm{0.5-2~keV}$ } & \colhead{$f_\mathrm{0.5-10~keV}$}  \\
\small{[mag]} & \small{[mag]} & \small{[erg cm$^{-2}$ s$^{-1}$]} & \small{[erg cm$^{-2}$ s$^{-1}$]}  \\
\hline
21.479 & 0.542 & $(4.5\pm0.6)$ & $(2.8\pm0.2)$ \\
$\pm0.008$ & $\pm0.011$ & $\times 10^{-15}$ & $\times 10^{-14}$\\ \hline \hline
\colhead{$M_\star$ } & \colhead{SFR} & \colhead{$\Psi$ } & \colhead{$L_\mathrm{AGN}$}  \\
\small{[$10^9~M_\odot$]} & \small{[$M_\odot$ yr$^{-1}$]} & & \small{[$10^{43}$ erg s$^{-1}$]}  \\
\hline
$3.0\pm0.7$ & $5.76\pm0.56$ & 40$^{\circ}$ & $9.56\pm0.48$ \\ \hline \hline
\colhead{$\log{\left(\frac{\mathrm{[OI]}}{\mathrm{H}\alpha}\right)}$} & \colhead{$\log{\left(\frac{\mathrm{[OIII]}}{\mathrm{H}\beta}\right)}$} & \colhead{$f_{\mathrm{H}\alpha}$} & \colhead{$f_\mathrm{[OIII]}$}  \\
 &  & \small{[erg cm$^{-2}$ s$^{-1}$]} & \small{[erg cm$^{-2}$ s$^{-1}$]} \\
\hline
-0.75 & 1.04 & $1.1\times 10^{-15}$ & $5.4\times 10^{-16}$ \\
\enddata
\end{deluxetable}

\section{Constraining the Stellar Mass} \label{sec:Mstar}

\begin{figure*}[htp]
    \centering
    \includegraphics[width=\linewidth]{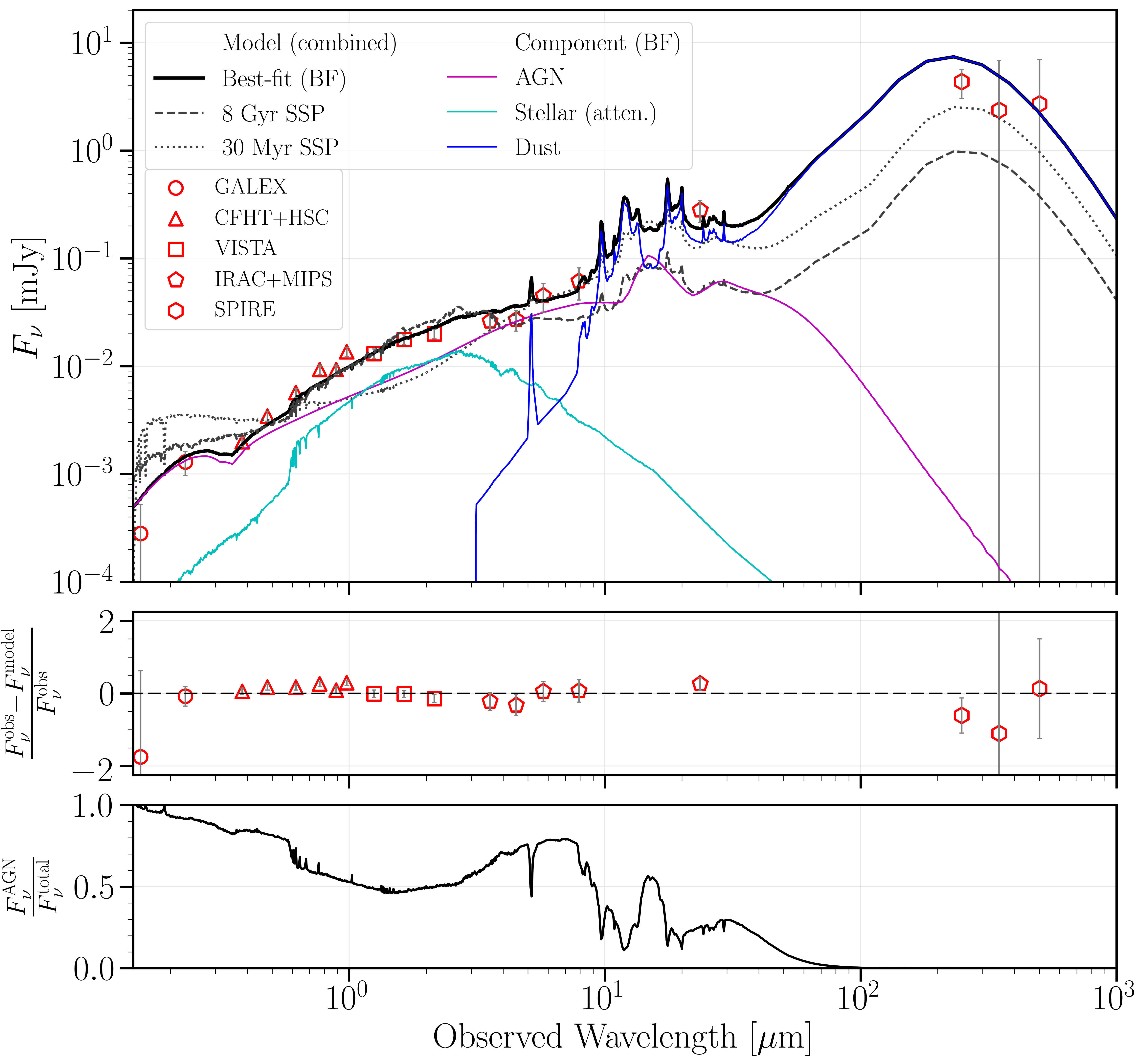}\caption{Result of SED modeling using CIGALE. Top panel: The best-fit model (solid black line) and two models with fixed single stellar populations (SSPs; gray lines) of ages 8 Gyr (dashed) and 30 Myr (dotted). Red symbols show photometric observations (which represent total magnitudes because the galaxy is barely resolved even in HSC) and relative residuals (middle panel) with shapes indicating different instruments or surveys. Errorbars represent $1\sigma$ statistical uncertainties with a 1\% systematic floor. Bottom panel: ratio of AGN to total emission in the best-fit model.}
    \label{fig:sed}
\end{figure*}

Our initial estimate of the stellar mass, $M_\star \approx 4 \times 10^9 M_\odot$, was derived from photometric template-fitting of the HSC $grizy$, CLAUDS $u^\star$, and GALEX data (see T.~Moutard et al., in preparation), following the procedure described by \citet{2016A&A...590A.103M}. We utilized additional photometry from {\it Spitzer} IRAC and MIPS, {\it Herschel} SPIRE \citep{2010A&A...518L...3G}, and the VISTA Deep Extragalactic Observations survey \citep[see][and references therein]{2018PASJ...70S...4A} to further constrain $M_\star$ using a multi-wavelength SED fit with the Code Investigating GALaxy Emission \citep[CIGALE \footnote{https://cigale.lam.fr/};][]{2009A&A...507.1793N,2019A&A...622A.103B}. In calculating the SED model, we used the stellar population models of \citet{2003MNRAS.344.1000B}, the \citet{2003PASP..115..763C} initial mass function, an exponentially declining delayed star-formation history, a dusty star-forming template from \citet{2014ApJ...784...83D}, and the AGN torus models of \citet{2006MNRAS.366..767F} allowing for a range of optical depths and inclinations. We fixed the extinction using the Balmer decrement from the H$\alpha$/H$\beta$ ratio ($A_V=1.3$; see \S \ref{sec:BH}), though we recover a similar $A_V$ when leaving it as a free parameter. The observed SED and CIGALE models are shown in Fig. \ref{fig:sed}. The best-fit model, for which we also show the residuals (middle panel), has $\chi^{2}_{n-1} = 2.87$ indicating a formally poor fit. However, this is expected because we exclude nebular/AGN emission lines in the templates. The photometric points that contribute most to raising the $\chi^2$ value are the HSC-$i$ and HSC-$Y$ bands, which fall precisely at the wavelengths including the high-equivalent width [OIII] and H$\alpha$ lines. These are under-estimated by our SED model.

The best-fit stellar model has $M_\star \approx (3.0\pm 0.7) \times 10^{9} M_\odot$, consistent with our initial estimate from CLAUDS, stellar population age $\sim0.5$~Gyr, and SFR $\approx (5.76\pm0.56)~ M_\odot~\mathrm{yr}^{-1}$. The AGN component dominates the rest-frame optical emission (see bottom panel of Fig. \ref{fig:sed}), favoring an intermediate inclination of $\Psi \approx 40^{\circ}$, placing HSC-XD 52 closer to the Type 1 regime (i.e. mostly unobscured).

To obtain a conservative $M_\star$ upper limit, we ran CIGALE with the same input ranges but required a fixed single stellar population (SSP) of 8 Gyr (the age of the Universe at $z\approx 0.5$). The upper limit of $M_\star < 1.1 \times 10^{10} M_\odot$ is shown as a gray dashed line in Fig. \ref{fig:sed}. Furthermore, we found a fixed SSP of 30 Myr (i.e., a very young stellar population; dotted gray line in Fig. \ref{fig:sed}) fails to reproduce the observed photometry.

Our case for the low stellar mass of the host galaxy is strengthened by spectral indicators of its low metallicity. In particular, both the SDSS spectrum and the higher resolution, higher SNR MagE spectrum show a complete lack of evidence for [NII] $\lambda$6548, $\lambda$6583 lines (see top panels of Fig. \ref{fig:fits}). We can put a limit on the line strength compared to H$\alpha$ from the MagE spectrum of [NII]/H$\alpha$ $\lesssim -1.8$. From the models of \citet{2006MNRAS.371.1559G}, we then place the metallicity at $\lesssim 0.25Z_\odot$, where $Z_\odot$ represents solar metallicity. Given the well-known stellar mass-metallicity relation, we can conclude by this entirely complementary line of evidence that HSC-XD 52 indeed qualifies as a low-mass galaxy.

\section{Evidence for a central black hole} \label{sec:BH}

\subsection{X-ray detection}
Combining observations from three separate epochs, with the first dominating the signal, HSC-XD 52 is strongly detected in the XMM soft (SB; $0.5-2$~keV), hard (HB; $2-10$~keV) and full (FB; $0.5-10$~keV) bands, with $\approx$ $109$, $133$, and $245$ photon counts (in the PN+MOS1+MOS2 detectors), respectively \citep{2018MNRAS.478.2132C}. It has a FB luminosity  of $L_X \approx 3.5 \times 10^{43}~\mathrm{erg}~\mathrm{s}^{-1}$ and an SNR of 185 (143) in the PN (M1) detector (see Table \ref{tab:props}). The derived hardness ratio of
\begin{equation}
    \mathrm{HR} \equiv \frac{H-S}{H+S} = 0.17\pm 0.07,
\end{equation}
where $H$ ($S$) is the total (all three detectors) net counts divided by the total exposure time in the HB (SB), is consistent with a Type I AGN at the redshift of HSC-XD 52. This XMM source is spatially coincident with the HSC optical source. Fig. \ref{fig:ims+spec} provides the XMM image with the X-ray and HSC centroids marked as a black cross and a red circle ($1.3''$), respectively.

X-ray emission produced by stellar processes, such as that from X-ray binaries (XRBs), could mimic the accretion signatures of AGN. From \citet{2010ApJ...724..559L}, the relation between XRB $L_X$, $M_\star$, and SFR is
\begin{equation}
    L_{X,\mathrm{XRB}} = (\alpha M_\star + \beta \mathrm{SFR})~ \mathrm{erg}~\mathrm{s}^{-1},
\end{equation}
with $\alpha = (9.05 \pm 0.37) \times 10^{28}~M_\odot^{-1}$ and $\beta = (1.62 \pm 0.22) \times 10^{39}~(M_\odot~\mathrm{yr}^{-1})^{-1}$. We have focused on high-mass XRBs because these dominate at $L_X \gtrsim 10^{39}~\mathrm{erg}~\mathrm{s}^{-1}$ \citep{2010ApJ...724..559L}. Our best estimates of $M_\star$ and SFR (Table \ref{tab:props}) generate an expected luminosity due to XRBs of $L_{X,\mathrm{XRB}} \approx 9.6 \times 10^{39}~\mathrm{erg}~\mathrm{s}^{-1}$, several orders of magnitude lower than that observed. Even adopting conservative upper limit values for SFR and $M_\star$ results in $L_{X,\mathrm{XRB}} \ll 10^{43}~\mathrm{erg}~\mathrm{s}^{-1}$.

\begin{figure}[htp]
    \centering
    \includegraphics[width=\linewidth]{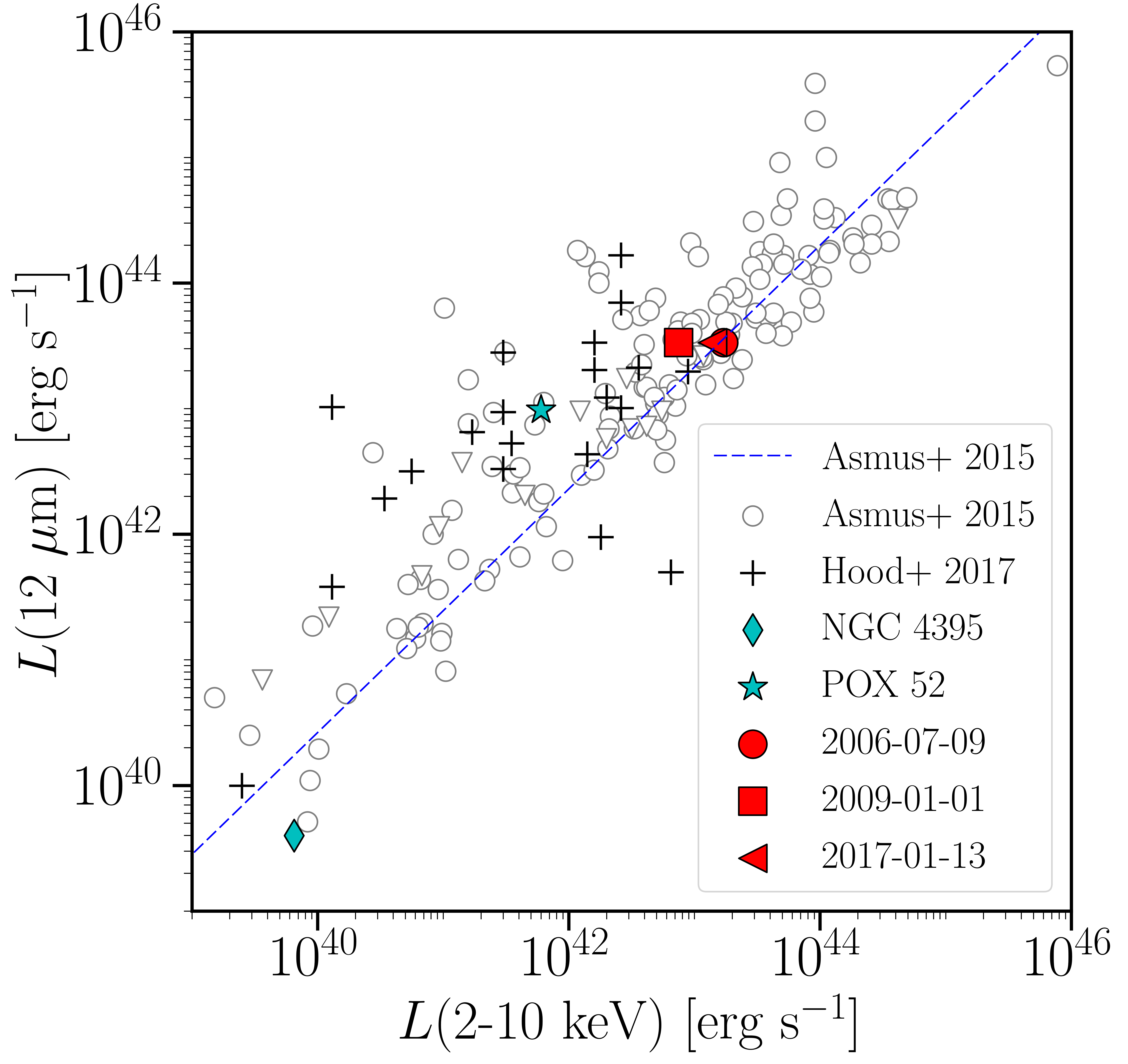}
    \caption{Observed correlation between IR luminosity at $\lambda = 12~\mu \mathrm{m}$ and X-ray luminosity at $2-10$ keV with the best-fit from \citet{2015MNRAS.454..766A} (dashed blue line). $L(12~\mu \mathrm{m})$ is not measured directly but extrapolated using a power-law fit to the IRAC and MIPS data. We include data compiled from \citet{2015MNRAS.454..766A} (all types of AGN; grey open symbols) and \citet{2017ApJ...838...26H} (low-mass AGN; black crosses). Triangles indicate upper limits. Cyan symbols represent low-redshift analogs to HSC-XD 52, which is shown at early, intermediate, and late epochs as a red circle, square, and triangle (upper limit on X-ray luminosity), respectively.}
    \label{fig:LIRLx}
\end{figure}

HSC-XD 52 was observed with XMM three separate times spanning over a decade: first on July 9 2006, next on January 1 2001, and finally on January 13 2017. The first two observations yielded fluxes (in cgs) of $1.35\pm0.28 \times 10^{-14}$ and $5.92\pm1.45 \times 10^{-15}$ from the PN detector alone. During the final epoch of observation, the source was detected only by the MOS2 detector because it fell on a dead chip of the MOS1 detector, and in a chip gap on the PN detector. Thus, this final observation yields only a weak upper limit on the flux (in cgs) of $<1.10 \times 10^{-14}$. In Fig. \ref{fig:LIRLx}, we compare HSC-XD 52 at the three different epochs to other X-ray luminous AGN observed with high spatial resolution ground-based mid-IR imaging \citep{2015MNRAS.454..766A} and low-mass AGN from \citet{2017ApJ...838...26H}. HSC-XD 52 falls on the empirical correlation (within the uncertainties) measured by \citet{2015MNRAS.454..766A} at least for the first two epochs. We also find the properties of HSC-XD 52 at these epochs to be consistent with other observed correlations (e.g. $L_X$--[OIII]), further bolstering our confidence in its characterization as an AGN. Comparing HSC-XD 52 with itself over the three epochs, it is clear that its X-ray luminosity is fading over time, suggesting a corresponding fall in activity.

To derive a lower limit on $M_{\rm BH}$, we can assume HSC-XD 52 is radiating at the Eddington luminosity when activity level is highest. We assume a bolometric luminosity as determined by the best-fit SED model of $L_\mathrm{AGN} = 1.20 \times 10^{44}~\mathrm{erg}~\mathrm{s}^{-1}$, which is also consistent with applying a typical bolometric correction to the observed $L_X$ at this earliest epoch. This yields a lower limit of
\begin{eqnarray}
    \nonumber M_\mathrm{BH} &\gtrsim \left(\frac{L_\mathrm{AGN}}{1.26 \times 10^{38}~\mathrm{erg}~\mathrm{s}^{-1}} \right) M_\odot \\ 
    &\gtrsim 9.5 \times 10^{5} M_\odot. \label{eq:Medd}
\end{eqnarray}

\subsection{Spectral diagnostics}

\begin{figure}[htp]
    \centering
    \includegraphics[height=3.7289cm]{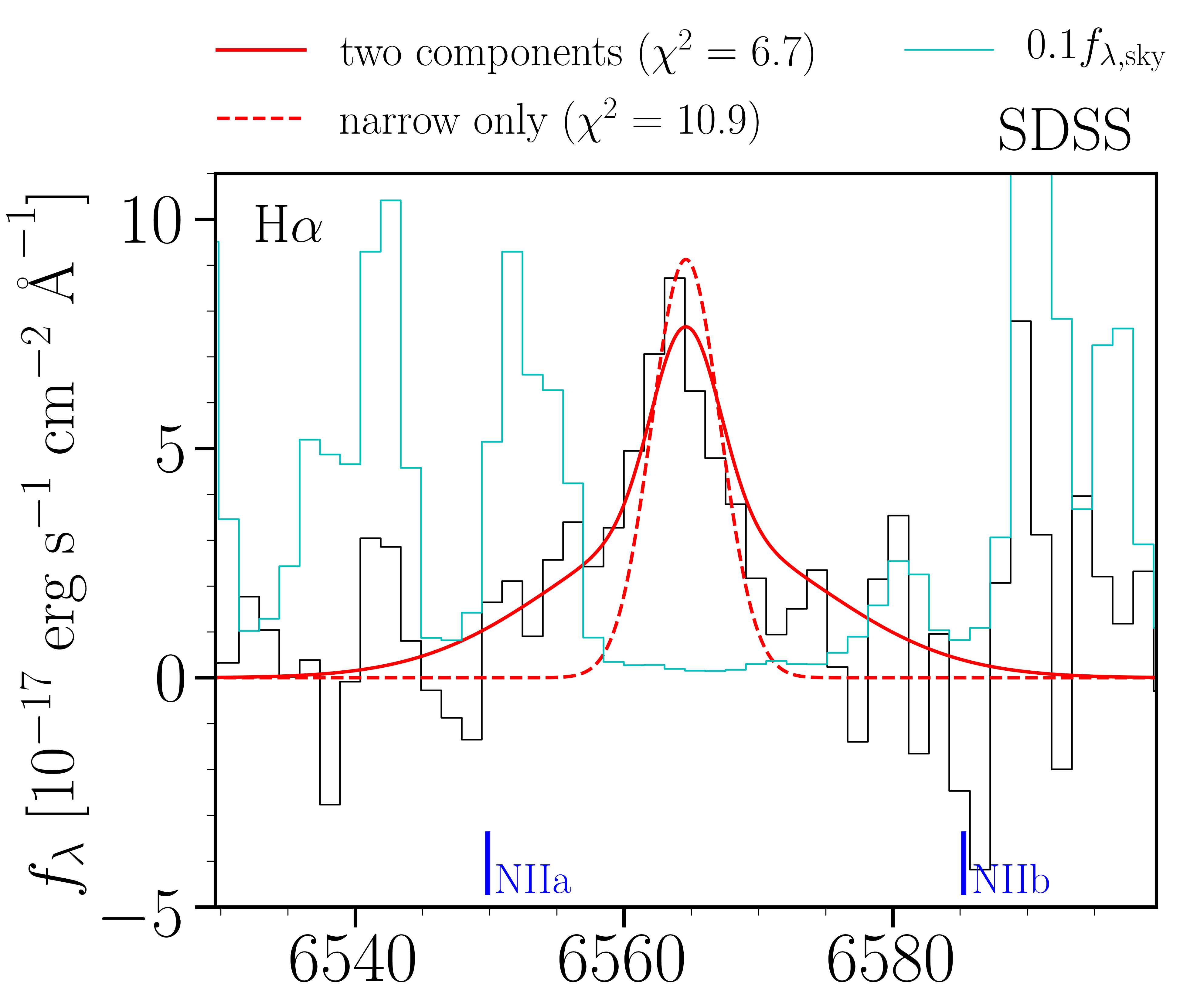}
    \includegraphics[height=3.7289cm]{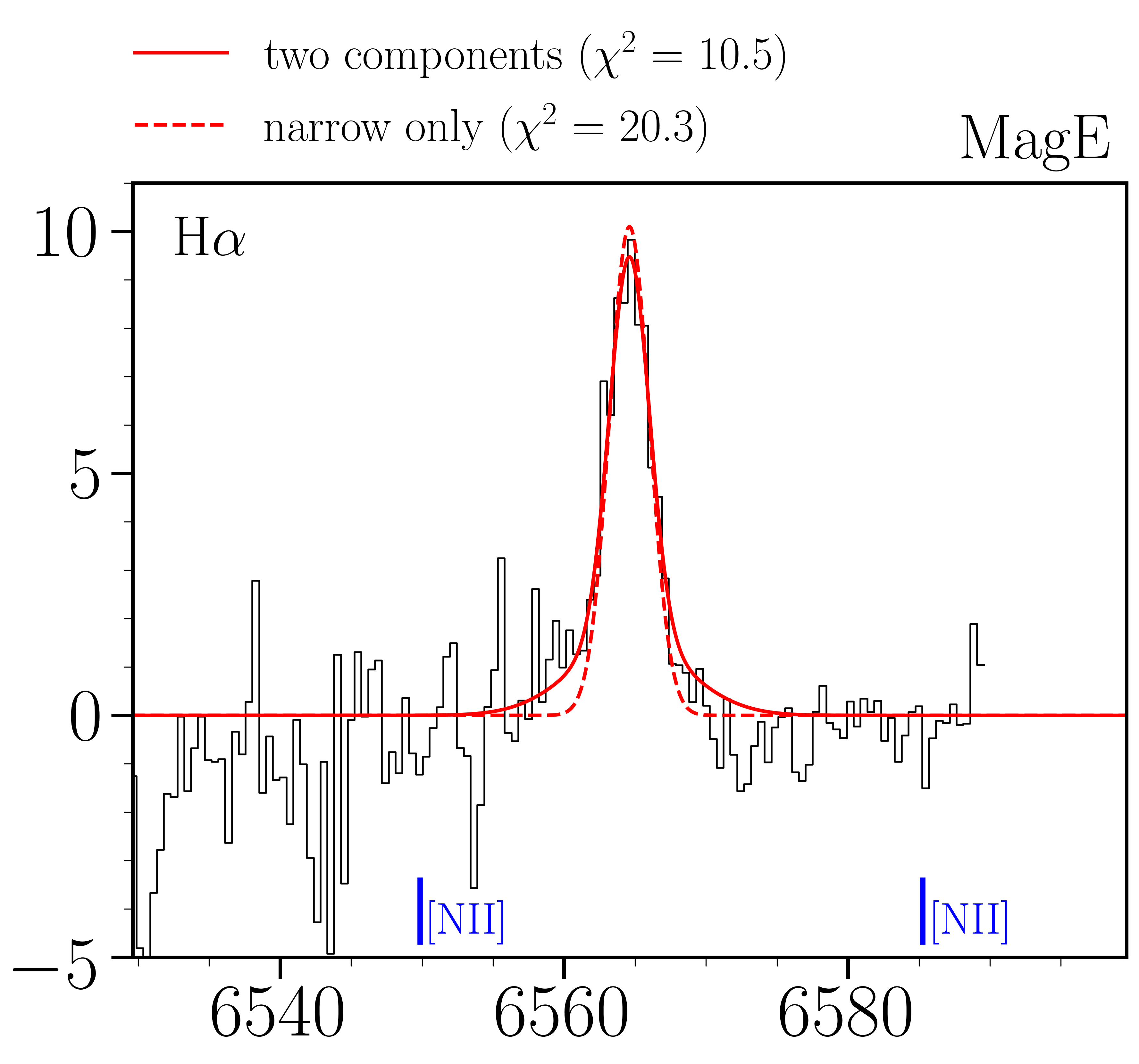}
    \includegraphics[width=\linewidth]{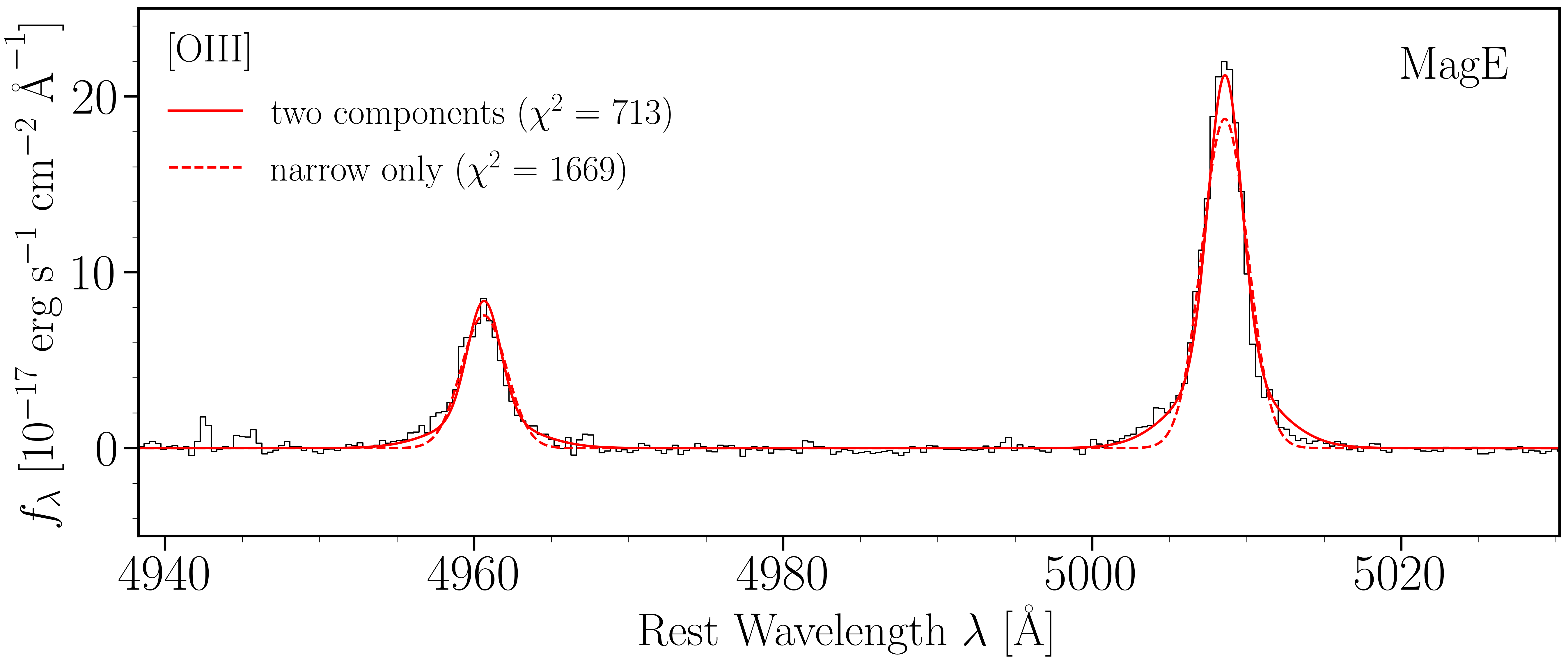}
    \caption{Zoom-ins of the SDSS (top left) and MagE (top right and bottom) spectra (black lines) with fitted models of important emission lines H$\alpha$ (top panels) and the [OIII 4959,5007] doublet (bottom). We model each line (pair of lines, in the case of [OIII] as a two-component Gaussian (solid red line) or as single narrow component Gaussian (dashed red line). In the H$\alpha$ zoom-ins (top), we also label where [NII] lines would fall if they were present. For the SDSS spectrum (top left), we overplot the sky spectrum scaled down by a factor of 10 (cyan line). None of these lines is highly contaminated, while H$\alpha$ in particular falls conveniently between sky lines.}
    \label{fig:fits}
\end{figure}

An SDSS spectrum from November 9 2013 is publicly available for this source. In addition, we acquired a spectrum with MagE on July 27 2019. The SDSS spectrum is dominated by strong [OIII] and Balmer emission lines. Our modeling of these lines, especially H$\alpha$ and H$\beta$, favors a broad-line component, suggesting the presence of an AGN contributing to the observed flux. However, the more recently acquired MagE spectrum, which has better resolution and higher SNR, does not exhibit broad lines. In Fig. \ref{fig:fits}, we show zoom-ins of the [OIII 4959,5007] doublet from the MagE spectrum (bottom panel) and the H$\alpha$ lines (top panel) from both the SDSS (left) and MagE (right) spectra. In red, we show our best-fit models for these lines, with dashed lines representing single-component Gaussians and solid lines representing composite Gaussians consistenting of both broad and narrow components. While the SDSS spectrum favors some broad component in the H$\alpha$ line, the MagE H$\alpha$ line is consistent with the [OIII] doublet. We note that the H$\beta$ emission in the MagE is blended with a strong sky line in the MagE spectrum, making it unreliable for an accurate line profile measurement.

Using the SDSS spectrum, for which we find broad H$\alpha$ and H$\beta$ alone, we measure a FWHM of $\mathrm{FWHM}_{\mathrm{H}\alpha} = \mathrm{FWHM}_{\mathrm{H}\beta}\approx 1076~\mathrm{km}~\mathrm{s}^{-1} $. From this spectrum, we also measure the continuum luminosity at $\lambda=5100$ \AA$~$of $L_{5100} \approx 1.8 \times 10^{43}~\mathrm{erg}~\mathrm{s}^{-1}$. Our SED modeling suggests that at this wavelength in the continuum, the AGN emission dominates compared to the starlight by a factor of $\approx 5$ (bottom panel of Fig. \ref{fig:sed}). We can then apply the virial formula presented as eqn. (5) of \citet{2005ApJ...630..122G} to estimate an effective upper limit on the black hole mass of $M_{\rm BH} \lesssim 1.7 \times 10^{6} M_\odot$. While this mass estimate is consistent with that based on the Eddington argument, we emphasize that it is based on a somewhat equivocal broad line detection.

As an additional verification of HSC-XD 52's AGN nature, we also measured the [OI 6300] emission line. The ratios of [OI 6300]/H$\alpha$ and [OIII 5007]/H$\beta$ allows us to place the source on an emission line diagnostic diagram, where it falls securely within the Seyfert region \citep[e.g.][]{2006MNRAS.372..961K} during the epoch of the SDSS spectrum.

\section{Discussion and conclusion} \label{sec:end}

We have presented photometric and spectroscopic observations of HSC-XD 52, an object identified as part of a new search for X-ray selected low-mass AGN in the HSC survey. Our analysis suggests that HSC-XD 52 has a stellar mass of $M_\star \approx (3.0\pm 0.7) \times 10^{9} M_\odot$ and hosts a luminous accreting black hole with $M_\mathrm{BH} \approx 10^{6} M_\odot$ and $L_X \approx 3.5 \times 10^{43}~\mathrm{erg}~\mathrm{s}^{-1}$, though its X-ray luminosity (and thus accretion rate) is variable and appears to be decreasing with time. Its properties resemble those of known low-mass AGN hosts in the local Universe, including POX 52 and NGC 4395. The detection of HSC-XD 52 provides convincing evidence for the existence of luminous active MBHs in low-mass galaxies at this higher redshift. 
Further observations, particularly temporal monitoring in rest-frame optical, could reveal continuum variability and thus shed more light on the evolution of AGN activity.

By requiring an unambiguous XMM detection, our methodology selects for the most luminous sources at intermediate redshifts. This is clear when we compare HSC-XD 52 to other, more local low-mass AGN including POX 52 and NGC 4395. As a direct consequence of our search method and the source's higher redshift ($z\approx 0.56$), HSC-XD 52 lies on the brighter end of observed correlations (e.g. between mid-IR and X-ray luminosities; Fig. \ref{fig:LIRLx}). Outside of the very local Universe, it is only for sources like this one that we can obtain a robust determination of the galaxy's low stellar mass, a confirmation of its AGN nature, and a well-constrained black hole mass. 

The best-fit model to the observed correlation between $M_\star$ and $M_\mathrm{BH}$ for local broad-line AGN of \citet{2015ApJ...813...82R} predicts $M_\mathrm{BH} \approx 7.1 \times 10^{5} M_\odot$ for HSC-XD 52, which is consistent with our estimate of the BH mass given the 0.55 dex scatter. With a larger sample of similar objects, we can begin to explore the relationship between $M_\mathrm{BH}$ and $M_\star$ for low-mass galaxies at intermediate redshifts in earnest.

To place meaningful constraints on statistical properties such as the occupation fraction as a function of redshift, we require a large sample with well-understood selection effects and biases, but significant challenges remain. While spectroscopic redshifts are cleaner, they impose a selection function on the samples used in past studies. Photometric redshifts are more inclusive, but often include undesirable contaminants as a repercussion, among other drawbacks (see details in G.~Halevi et al. 2019, in preparation).  Furthermore, probing the regime of lower luminosity X-ray sources requires unambiguously discriminating between stellar processes (i.e. XRBs) and accretion onto an MBH as the origin of the X-rays. This becomes increasingly difficult at $L_X \lesssim 10^{40}~\mathrm{erg}~\mathrm{s}^{-1}$, which presents a significant obstacle to developing a clean sample of low-mass AGN that is representative of the underlying population while still covering a sufficiently large volume.

In the 2030s, the launch of the next-generation X-ray observatory Lynx \citep{2018arXiv180909642T} would enable high SNR X-ray spectra of AGN in dwarf galaxies, and with its high spatial resolution, localization of the X-ray emission, thus breaking the degeneracy between emission from AGN and stellar processes. Additionally, its sensitivity will facilitate X-ray detections of SMBH seeds at cosmological distances. Meanwhile, the space-based gravitational wave detector LISA \citep{2017arXiv170200786A} will open the door to multi-messenger studies of early SMBHs. These future missions imply a promising future for the quest to illuminate the mysteries of how MBHs form and grow, and in turn, how their host galaxies evolve.

\acknowledgements
We thank the anonymous referee for their helpful comments. The HSC collaboration includes the astronomical communities of Japan and Taiwan, and Princeton University. The HSC instrumentation and software were developed by the National Astronomical Observatory of Japan (NAOJ), the Kavli Institute for the Physics and Mathematics of the Universe (Kavli IPMU), the University of Tokyo, the High Energy Accelerator Research Organization (KEK), the Academia Sinica Institute for Astronomy and Astrophysics in Taiwan (ASIAA), and Princeton University. Funding was contributed by the FIRST program from Japanese Cabinet Office, the Ministry of Education, Culture, Sports, Science and Technology (MEXT), the Japan Society for the Promotion of Science (JSPS), Japan Science and Technology Agency (JST), the Toray Science Foundation, NAOJ, Kavli IPMU, KEK, ASIAA, and Princeton University. 
This work is based on observations obtained with MegaPrime/MegaCam, a joint project of CFHT and CEA/DAPNIA, at the Canada-France-Hawaii Telescope (CFHT) which is operated by the National Research Council (NRC) of Canada, the Institut National des Science de l'Univers of the Centre National de la Recherche Scientifique (CNRS) of France, and the University of Hawaii. This research uses data obtained through the Telescope Access Program (TAP), which has been funded by the National Astronomical Observatories, Chinese Academy of Sciences, and the Special Fund for Astronomy from the Ministry of Finance. This work uses data products from TERAPIX and the Canadian Astronomy Data Centre. It was carried out using resources from Compute Canada and Canadian Advanced Network For Astrophysical Research (CANFAR) infrastructure. 
Based in part on data products from observations made with ESO Telescopes at the La Silla Paranal Observatory as part	of the VISTA Deep Extragalactic	Observations \citep[VIDEO][]{2013MNRAS.428.1281J} survey, under program ID 179.A-2006 (PI: Jarvis).
Support for the design and construction of the Magellan Echellette 
Specrograph was received from the Observatories of the Carnegie Institution of 
Washington, the School of Science of the Massachusetts Institute of Technology, and 
the National Science Foundation in the form of a collaborative Major 
Research Instrument grant to Carnegie and MIT (AST0215989).
Based in part on observations made with the Spitzer Space Telescope, which is operated by the Jet Propulsion Laboratory, California Institute of Technology under a contract with NASA.
Funding for SDSS IV has been provided by the Alfred P. Sloan Foundation, the U.S. Department of Energy Office of Science, and the Participating Institutions. 
We recognize the cultural role and reverence of the Maunakea summit within the indigenous Hawaiian community; we are grateful for the access that enables observations crucial to this work.
\textit{Software:} \texttt{Astropy} \footnote{http://www.astropy.org} \citep{2013A&A...558A..33A,2018AJ....156..123A}

\bibliography{dwarfAGN.bib}

\end{document}